\documentclass[a4paper,prb]{revtex4}
\usepackage{amsmath}
\usepackage{graphicx}
\usepackage{color}
\usepackage{subfigure}
\usepackage{hyperref}
\begin{document}
\title{Left Passage Probability of SLE$(\kappa,\rho)$}
\author {M. N. Najafi }
\affiliation{Physics department, Sharif University of Technology, P.O. Box 11155-9161, Tehran, Iran}

\begin{abstract}
SLE($\kappa,\vec{\rho}$) is a variant of the Schramm-Loewner Evolution which describes the curves which are not conformal invariant, but are self-similar due to the presence of some other preferred points on the boundary. In this paper we study the left passage probability (LPP) for SLE($\kappa,\vec{\rho}$) through field theoretical framework and find the differential equation which govern this probability. This equation is solved (up to two undetermined constants) for the special case $\kappa=2$ and $h_{\rho}=0$ for large $x_0$ at which the boundary condition changes. This case may be referred to the Abelian sandpile model with a sink on the boundary. As an example, we apply this formalism to SLE($\kappa,\kappa-6$) which governs the curves that start from and end on the real axis.
\end{abstract}
\maketitle
\section{Introduction}

    The recent breakthrough in 2-dimensional (2D) critical phenomena, referred to Schramm-Loewner Evolution (SLE), has provided us with a new interpretation of the traditional conformal field theory (CFT) and Coulomb gas approaches. According to Schramm's idea \cite{Shramm} one can describe the interfaces of 2D critical statistical models via a stochastic growth processes in which statistical models fall into one-parameter classes labelled by a diffusivity parameter, namely $\kappa$. The examples of statistical models which are described by SLE are Ising model \cite{Smir}, Potts model \cite{Potts}, $O(n)$ model \cite{O(n)}, Abelian sandpile model (ASM) \cite{Saberi} etc. and some geometrical models such as self avoiding walks \cite{SAW}, percolation \cite{Smir2}, loop erased random walk (LERW) \cite{LERW} etc. This description focuses on some non-local objects in the statistical models, in contrast to  CFT in which one deals with local fields. These non-local objects can be interfaces of statistical models, be it the boundary of clusters in Ising model, or the boundary of avalanches in ASM, or loops in $O(n)$ model. \\
    A very crucial step towards understanding SLE was taken by D. Bernard et. al. \cite{BauBer} to connect this theory to CFT. They found a simple relation between the diffusivity parameter $\kappa$ in SLE and the central charge $c$ in CFT. This connection helps to conjecture a CFT universality class for a less-known statistical model, having its diffusivity parameter (numerically or theoretically) and vice versa. One generally hopes to have some operators in the CFT content, corresponding to each SLE-observable in the statistical model in hand with a specific $\kappa$. The examples are crossing probability and left passage probability (LPP) \cite{Schramm2,Kytola}. The inverse of this statement is also true, i.e. statistical observables (such as LPP) can be used to obtain the diffusivity parameter of the model in hand. In fact the most reliable methods to obtain $\kappa$ for a generic critical statistical model lies within these analysis's such as the winding angle distribution and LPP of the SLE curves. LPP of chordal SLE can be expressed in terms of $\kappa$ within Schramm's formula \cite{BerBau}. This probability is the solution of a differential equation obtained by conformal invariance of the probability measure of the growing SLE curve.
 
SLE($\kappa,\vec{\rho}$) is a variant of SLE($\kappa$) in which there are some more preferred points on the boundary, affecting the growth process of the SLE curve. The relation of this generalization of SLE to CFT and its operator content and also its correspondence to the Coulomb gas is widely studied \cite{Cardy-clmp,Cardy,BerBau-rev}. Since in some models, we deal with the interfaces which have these preferred points at the boundary of their domain, the calculation of the statistical observables for SLE($\kappa,\rho$) seems to be crucial in calculating the diffusivity parameter of the them. As an example, one can mention the statistics of the avalanche frontiers, defined in Abelian sandpile model (ASM) in presence of a sink point in which the grains dissipate. It is numerically known that these avalanche frontiers, when there is no dissipation, are SLE($\kappa=2$) \cite{Saberi}. When in some point on the boundary the grains dissipate, the statistics of these frontiers change so that they are not necessarily ordinary SLE and may be analyzed within other variants of SLE. In this paper we analyze the LPP of such curves with some preferred points on the boundary which are described by SLE($\kappa,\vec{\rho}$) and especially present the results for the case mentioned above i.e. Abelian sandpile model (ASM) in presence of a sink point.

In the next section we briefly introduce SLE and its variant SLE($\kappa,\vec{\rho}$). Sections \ref{LPP} and \ref{LPP2} are devoted to the LPP of the SLE($\kappa,\kappa-6$) and the more general case SLE($\kappa,\vec{\rho}$). In section \ref{num} we present the results for the case $\kappa=2$ and $h_{\rho}=0$ for large $x_0$ at which the boundary condition changes.

\section{SLE}\label{SLE}
SLE theory describes the critical behavior of 2D statistical models by focusing on their geometrical features such as their interfaces. These domain walls are some non-intersecting curves which directly reflect the status of the system in question and are supposed to have two properties: conformal invariance and the domain Markov property. SLE is the candidate to analyze these random curves by classifying them to the one-parameter classes SLE$_{\kappa}$. For good introductory review see references \cite{RohdeSchramm,Cardy}. There are three kinds of SLE; chordal SLE in which the random curve starts from zero and ends at infinity, dipolar SLE in which the curve starts from and ends at the boundary and radial SLE in which the curve starts from the boundary and ends in the bulk. In this paper we deal with chordal and dipolar SLEs.
\subsection{Chordal SLE}
Let us denote the upper half-plane by $H$ and $\gamma_t$ as the SLE trace grown up to time $t$. SLE$_{\kappa}$ is a growth process defined via conformal maps which are solutions of stochastic Loewner's equation:
\begin{equation}
\partial_{t}g_{t}(z)=\frac{2}{g_{t}(z)-\xi_{t}},
\label{Loewner}
\end{equation}
in which the initial condition is $g_{t}(z)=z$  and the driving function $\xi_{t}$ is proportional to a one dimensional Brownian motion $B_{t}$ i.e. $\xi_{t}=\sqrt{\kappa}B_{t}$ in which $\kappa$ is the diffusivity parameter defined above. $\tau_{z}$ is defined as the time for which for fixed $z$, $g_{t}(z)=\xi_{t}$ and the hull as $K_{t}=\overline{\lbrace z\in H:\tau_{z}\leq t \rbrace}$. It is notable that the complement $H_{t}:=H\backslash{K_{t}}$ is simply-connected so that one can conclude that every point which is separated from the infinity by the SLE trace will be involved in $K_t$. The map $g_{t}(z)$ is well-defined up to time $\tau_z$. This map is the unique conformal mapping $H_{t}\rightarrow{H}$ with $g_{t}(z)=z+\frac{2t}{z}+O(\frac{1}{z^{2}})$ as $z\rightarrow{\infty}$ known as hydrodynamical normalization. 

There are three phases for SLE traces; for $0<\kappa\leq{4}$ the trace is non-self-intersecting and it does not hit the real axis;  in this case the hull and the trace are identical: $K_{t}=\gamma_{t}$. This is called "dilute phase". For $4<\kappa<8$, the trace touches itself and the real axis so that a typical point is surely swallowed as $t\rightarrow\infty$ and $K_{t}\neq\gamma_{t}$. This phase is called "dense phase". Finally for $\kappa\geq 8$ the trace is space filling. There is a connection between the first two phases: for $4\leq\kappa\leq{8}$ the frontier of $K_{t}$, i.e. the boundary of $H_{t}$ minus any portions of the real axis, is a simple curve which is locally a SLE$_{\tilde{\kappa}}$ curve with $\tilde{\kappa}=\frac{16}{\kappa}$, i.e. it is in the dilute phase \cite{Dub}.
The crucial question about the connection between SLE and CFT has been addressed by M. Bauer et.al. \cite{BauBer} in which it was shown that the bcc operator in CFT corresponding to the change of boundary condition at the point from which the SLE trace starts or ends, is the operator having null vector at second level with conformal weight $h_1(\kappa)=\frac{6-\kappa}{2\kappa}$ and the central charge $c=\frac{(3\kappa-8)(6-\kappa)}{2\kappa}$. This observation helps us to construct the CFT correspondence of the observables in SLE as we will see in the following sections.

\subsection{SLE($\kappa,\vec{\rho}$)}\label{SLE(k,r)}
As above, we define SLE($\kappa,\vec{\rho}$) in the upper half plane. The parameter $\kappa$, as was defined above, identifies the local properties of the model in hand, and the parameters $\vec{\rho}\equiv(\rho_1,\rho_2,...,\rho_n)$ has to do with the boundary condition changes (bc) imposed on the points on the real axis $x_1$,$x_2$,...,$x_n$ (except the origin from which the curve starts). For example, in the radial set up of SLE (with no further changes on the boundary), we have $n=1$ and $\rho=\kappa-6$ and the planar curves start from the origin and end on one point on the real axis (we name it as $x_{\infty}$)\cite{BerBau}. The stochastic equation governing such curves is the same as formula (\ref{Loewner}) but the driving function has a different form: 
\begin{equation}
d\xi_{t}=\sqrt{\kappa}dB_{t}+\frac{\rho_1}{\xi_{t}-g_{t}(x_1)}dt+\frac{\rho_2}{\xi_{t}-g_{t}(x_2)}dt+...+\frac{\rho_n}{\xi_{t}-g_{t}(x_n)}dt
\label{SLE(k,r)driving}
\end{equation}
The example is the radial SLE in which the driving function obeys the following equation\cite{BauBer2}:
\begin{equation}
d\xi_{t}=\sqrt{\kappa}dB_{t}+\frac{\kappa-6}{\xi_{t}-g_{t}(x_{\infty})}dt
\label{driving}
\end{equation}
Thus for the radial SLE (from real axis to itself), or for chordal case with some other preferred points on the real axis, the corresponding driving function acquires a drift term. For review see the references\cite{RohdeSchramm,Cardy}.

In determining the operator content of CFT of these models, the important feature is that the conformal weight of the bcc operators of each preferred point is obtained from a simple relation i.e. $h_{\rho}(\kappa)=\frac{\rho(\rho+4-\kappa)}{4\kappa}$\cite{Kytola}. We will use this in the following sections in order to determine the conformal weight of the probabilities.

\section{An Example: LPP of SLE($\kappa,\rho_c(=\kappa-6)$)}\label{LPP}

In this section we consider the probability of the event that a point $(x,y)$ lie to the right of a SLE curve i.e. $P(w,\bar{w},\xi_0,x^\infty)$ in which $w=x+iy$ is the detection point and $\bar{w}$ is its complex conjugate and $\xi_0$ is the point from which the SLE trace starts and $x^{\infty}$ is the point at which the curve ends. This probability has been calculated for the chordal case\cite{}. The Fig [\ref{sample}] schematically shows the situation. To proceed, we introduce the coordinates $\theta$ and $\phi$, indicated in the figure as follows:

\begin{figure}
\centerline{\includegraphics[scale=.45]{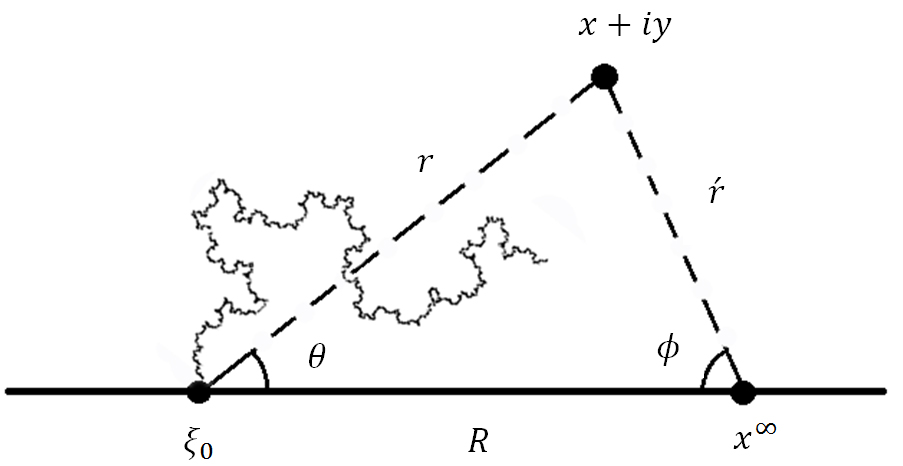}}
\caption{The schematic picture of a triangle involving the points $\xi_0$, $x^\infty$ and $x+iy$.}
\label{sample}
\end{figure}

\begin{equation}
\begin{split}
&z-\xi_0 =re^{i\theta}=R\frac{\sin\phi}{\sin(\phi+\theta)}e^{i\theta}\\
&z-x^\infty =re^{i(\pi -\phi)}=R\frac{\sin\theta}{\sin(\phi+\theta)}e^{-i\phi}\\
&x^\infty-\xi_0 =R\\
&z-\bar{z}=R\frac{\sin\phi}{\sin(\phi+\theta)}\left(r^{i\theta}-r^{-i\theta} \right) 
\end{split}
\label{teta_phi}
\end{equation}
The use of these coordinates will facilitate the equations, as we will see in following.
\subsection{Calculation with SLE}
In this section we try to obtain the differential equation governing the probability $P(w,\bar{w},\xi_0,x^\infty)$ corresponding to (the LPP of) SLE($\kappa,\rho_c$) by using its invariance under conformal mappins. In this case, we should use the Eq [\ref{driving}] to obtain the change of $P(w,\bar{w},\xi_0,x^\infty)$ under the evolution of SLE trace and demand that (the ensemble average of) this function be invariant under such an operation: 
\begin{equation}
P(w,\bar{w},\xi_0,x^\infty)\longrightarrow \textbf{E} \left[ P(w+\frac{2dt}{w-\xi_0},\bar{w}+\frac{2dt}{\bar{w}-\xi_0},\xi_0+d\xi_0,x^\infty+\frac{2dt}{x^\infty -\xi_0}) \right] 
\end{equation}
in which $w=x+iy$ and $\bar{w}$ is its complex conjugate and $\textbf{E}$[ ] is the ensemble average. It is not difficult to obtain the final differential equation resulting from the Ito calculations:

\begin{equation}
\begin{split}
\left[ \frac{2}{w-\xi_0}\partial_w +\frac{2}{\bar{w}-\xi_0}\partial_{\bar{w}}+\rho_c \frac{1}{\xi_0-x^\infty}\partial_{\xi_0}-\frac{2}{\xi_0-x^\infty}\partial_{x^\infty}+\frac{\kappa}{2}\partial_{\xi_0}^2 \right] P(w,\bar{w},\xi_0,x^\infty)=0
\end{split}
\label{final_LPE}
\end{equation}
It is obvious that this equation has the symmetry under transformation $\xi_0\rightarrow \xi_0+a$ , Re[$w$]$\rightarrow$ Re[$w$]$+a$ , $x^\infty\rightarrow x^\infty+a$. So one can easily check that in this equation $\partial_{x^\infty}=-\partial_{\xi_0}-\partial_x$, ($x=$Re[$w$]). To proceed, we need to predict the form of $P(w,\bar{w},\xi_0,x^\infty)$. In the next subsection we will see that using conformal field theory (CFT), we can reduce the Eq [\ref{final_LPE}] to a single-variable differential equation and relate it to the chordal one.
\subsection{CFT Background}

In the previous subsection, we have introduced the equation governing left passage probability. As this equation depend on three independent variable, it seems difficult to solve it. In this section we study the CFT interpretation of this quantity. Suppose that $\hat{O}$ is an operator which detects the left passage of SLE trace, i.e. the LPP is the expectation value of this operator in CFT content to the underling model. As we have boundary conformal field theory (real axis) with two boundary changing operators (one in $\xi_0$ and another in $x^\infty$), the LPP can be written as:

\begin{equation}
P(w,\bar{w},\xi_0,x^\infty)=\frac{\left\langle \hat{O}(x,y)\hat{O}(x,-y)\psi(\xi_0)\psi(x^\infty)\right\rangle}{\left\langle \psi(\xi_0)\psi(x^\infty)\right\rangle}
\label{corr}
\end{equation}
In this equation, the operator $\hat{O}(x,-y)$ is the image of $\hat{O}(x,y)$ with respect to the real axis. $\psi$ is the boundary changing operator for the CFT corresponding to the underling SLE whose conformal weight is $h_1(\kappa)=\frac{6-\kappa}{2\kappa}$ with second level null vector:
\begin{equation}
\left( \frac{\kappa}{2}L_{-1}^2-2L_{-2}\right) \psi=0
\end{equation}
leading to the following equation for the corresponding correlation function:
\begin{equation}
\left( \frac{\kappa}{2}\mathcal{L}_{-1}^2-2\mathcal{L}_{-2}\right)f(w,\bar{w},\xi_0,x^\infty)=0
\end{equation}
with
\begin{equation}
\begin{split}
f(w,\bar{w},\xi_0,x^\infty)&\equiv\left\langle \hat{O}(x,y)\hat{O}(x,-y)\psi(\xi_0)\psi(x^\infty)\right\rangle\\
&=\frac{1}{\left( \xi_0-x^\infty\right)^{2h_1(\kappa)}}P(w,\bar{w},\xi_0,x^\infty)\\
\mathcal{L}_{-1}&=\frac{\partial }{\partial \xi_0}\\
\mathcal{L}_{-2}&=\sum_i \left( \frac{h_i}{(z_i-\xi_0)^2}-\frac{1}{z_i-\xi_0} \frac{\partial}{\partial z_i} \right).
\end{split}
\label{f_dif}
\end{equation}
The above sum is over each field in the correlation function [\ref{corr}] except $\xi_0$. Substituting Eq [\ref{f_dif}] in Eq [\ref{corr}] yields the equation:
\begin{equation}
\left[ \frac{\kappa}{4}\partial_{\xi_0}^2- h_O\textrm{Re}\left( \frac{1}{(z-\xi_0)^2} \right)+\frac{1}{z-\xi_0} \frac{\partial}{\partial z}+\frac{1}{\bar{z}-\xi_0} \frac{\partial}{\partial \bar{z}}+\frac{1}{x^\infty-\xi_0} \frac{\partial}{\partial x^\infty}- \frac{h_1(\kappa)}{(x^\infty-\xi_0)^2}\right] f=0.
\label{LPE}
\end{equation}
In this equation $h_O$ is conformal weight of $\hat{O}$. It is notable that this equation can be written in terms of $x^\infty$ in which one exchanges the rule of $\xi_0$ and $x^\infty$. The chordal case can be obtained in the limit $x^\infty\rightarrow \infty$. In this limit we have $f=P(w,\bar{w},\xi_0,x^\infty)$ in Eq [\ref{f_dif}] and:
\begin{equation}
\left[ \frac{\kappa}{4}\partial_{\xi_0}^2- h_O\textrm{Re}\left( \frac{1}{(z-\xi_0)^2} \right)+\frac{1}{z-\xi_0} \frac{\partial}{\partial z}+\frac{1}{\bar{z}-\xi_0} \frac{\partial}{\partial \bar{z}}\right] f=0.
\label{LPE2}
\end{equation}
Comparing the Eq [\ref{LPE}] with the equation of LPP of the chordal case \cite{Cardy}, one obtains $h_O=0$. Substituting $f$ from Eq [\ref{f_dif}] into Eq [\ref{LPE}], it easy to check that the equation governing $P$ is the same as Eq [\ref{final_LPE}]. It is known that the global conformal symmetry can fix four point functions up to a function of the crossing ratios \cite{Dif}. In this case, letting $h_O=0$ we have:
\begin{equation}
\begin{split}
P&=y^{\frac{2}{3}h_1(\kappa)}(x^\infty-\xi_0)^{\frac{2}{3}h_1(\kappa)}[(x-x^\infty)^2+y^2]^{\frac{-1}{3}h_1(\kappa)}[(x-\xi_0)^2+y^2]^{\frac{-1}{3}h_1(\kappa)}g(\kappa,\eta,\bar{\eta})\\
&=\left( \eta\bar{\eta}\right)^{\frac{-1}{3}h_1(\kappa)} g(\kappa,\eta,\bar{\eta})\equiv \frac{1}{2}\left( \eta+\bar{\eta} \right) h(\kappa,\eta,\bar{\eta}).
\end{split}
\label{FPF}
\end{equation}
In the above formula $\eta$ is the crossing ratio i.e. $\frac{(w-\xi_0)(\bar{w}-x^\infty)}{y(x^\infty-\xi_0)}$ and $\bar{\eta}$ is its complex conjugate. So the finding of $P$ reduces to finding $h$. Let $u\equiv$Re$[\eta]=\frac{x(x-x^{\infty})+y^2}{yx^{\infty}}$ (we set $\xi_0=0$). After some calculations one obtains:
\begin{equation}
4u\partial_uP+\frac{\kappa}{2}\left( u^2+1\right) \partial_u^2P=0
\label{EFPF}
\end{equation}
In terms of $\theta$ and $\phi$, $u$ is equal to $\cot(\theta+\phi)$. The solution of this equation, with the boundary conditions $P=1$ for $\theta=\pi$ and $P=0$ for $\theta=0$ is:

\begin{equation}
P=\frac{1}{2}+\frac{\Gamma(\frac{4}{\kappa})}{\sqrt{\pi}\Gamma(\frac{8-\kappa}{2\kappa})} {_2F_1}\left(\frac{1}{2},\frac{4}{\kappa},\frac{3}{2},-\cot^2(\theta+\phi) \right)\cot(\theta+\phi) 
\label{EFPF2}
\end{equation}
This result can be derived directly from chordal case (the $x^{\infty}\rightarrow\infty$ limit, or equivalently $\phi\rightarrow0$). In this case it has been proved that \cite{Cardy}:
\begin{equation}
P_{\text{chordal}}=\frac{1}{2}+\frac{\Gamma(\frac{4}{\kappa})}{\sqrt{\pi}\Gamma(\frac{8-\kappa}{2\kappa})} {_2F_1}\left(\frac{1}{2},\frac{4}{\kappa},\frac{3}{2},-\left(  \frac{x-\xi_0}{y}\right) ^2 \right)\frac{x-\xi_0}{y} 
\end{equation}
The corresponding probability for the dipolar SLE can be obtained using the map $\varphi=\frac{x^{\infty}w}{(x^{\infty}-w)}$. Under this map, $x+iy\rightarrow\frac{x^{\infty}(x x^{\infty}-x^2-y^2)}{(x-x^{\infty})^2+y^2}+i\frac{x_{\infty}^2y}{(x-x^{\infty})^2+y^2}$. The probability that the point $x+iy$ is swallowed by a SLE curve in the dipolar case, is equal to the probability of left passage of the same point in the chordal set up i.e. the left passage probability of the mapped point $\varphi(x+iy)$. Using this point, one can write:
\begin{equation}
\begin{split}
P_{\text{dipolar}}(x+iy)&=P_{\text{chordal}}(\varphi(x+iy))\\
&=\frac{1}{2}+\frac{\Gamma(\frac{4}{\kappa})}{\sqrt{\pi}\Gamma(\frac{8-\kappa}{2\kappa})} {_2F_1}\left(\frac{1}{2},\frac{4}{\kappa},\frac{3}{2},-\left(  \frac{(x-\xi_0)(x^{\infty}-x)-y^2}{y(x^{\infty}-\xi_0)}\right) ^2 \right)\frac{(x-\xi_0)(x^{\infty}-x)-y^2}{y(x^{\infty}-\xi_0)} 
\end{split}
\end{equation}
which is exactly the same as Eq [\ref{EFPF2}] (setting $\xi_0=0$).

\section{LPP of SLE($\kappa,\rho$)}\label{LPP2}

In this section we apply the CFT formalism developed in the previous section to the more general case SLE($\kappa,\rho$). Lets consider a curve growing from origin to infinity, conditioned by a change in value of fields on the boundary which correspond to a scaling operator on this point with the weight $h_\rho=\frac{\rho(\rho+4-\kappa)}{4\kappa}$ (it is proved that this curve is a SLE($\kappa,\rho$))\cite{Kytola}. So the left passage probability equals to a five point function in the corresponding conformal field theory:

\begin{equation}
P(x,y,\xi_0,x_0)=\frac{\left\langle \hat{O}(x,y)\hat{O}(x,-y)\psi(x_0)\psi(\xi_0)\psi(\infty)\right\rangle }{\left\langle \psi(x_0)\psi(\xi_0)\psi(\infty)\right\rangle}
\label{LPP-CFT}
\end{equation}
In this regard, the problem reduces to the calculation of 3-point and 5-point functions which satisfy the boundary conditions. As above, we define $f(x,y,\xi_0,x_0)$ the numerator of the r.h.s. of Eq [\ref{LPP-CFT}]. So we have:

\begin{equation}
P(x,y,\xi_0,x_0)=(x_0-\xi_0)^{h_\rho}f(x,y,\xi_0,x_0)
\label{LPP-CFT2}
\end{equation}
Using the null vector equation for $\psi$ and after some calculations we obtain the following equation for $f$:

\begin{equation}
\left\lbrace \frac{\kappa}{2}\partial_{\xi_0}^2+\frac{2}{z-\xi_0}\partial+\frac{2}{\bar{z}-\xi_0}\bar{\partial}+\frac{2}{x_0-\xi_0}\partial_{x_0}+\frac{\kappa h_{\rho}}{x_0-\xi_0}\partial_{\xi_0}+\frac{\frac{1}{2}h_{\rho}[\kappa(h_{\rho}+1)-8]}{(x_0-\xi_0)^2}\right\rbrace  P=0
\label{LPP-SLE(kr)}
\end{equation}
Using global conformal invariance, one can fix $f$ up to a function of crossing ratios and prove that:

\begin{equation}
P=\left[ \left( x-\xi_0\right)^2+y^2\right]^{-\frac{1}{3}h_1+\frac{1}{6}h_{\rho}} \left[ \left( x-x_0\right)^2+y^2 \right]^{\frac{1}{3}h_1-\frac{1}{2}h_{\rho}} \left( x_0-\xi_0\right)^{-\frac{1}{3}h_1+\frac{1}{2}h_{\rho}}y^{\frac{1}{3}h_1+\frac{1}{6}h_{\rho}}g(\eta_1,\eta_2)
\label{LPP-CFT3}
\end{equation}
in which we have considered two independent crossing ratios $\eta_1\equiv \frac{(z-\xi_0)(\bar{z}-x_0)}{y(x_0-\xi_0)}$ and $\eta_2\equiv \frac{(z-\xi_0)}{y}=\lim_{x^{\infty}\rightarrow\infty}\frac{(z-\xi_0)(\bar{z}-x^{\infty})}{y(x^{\infty}-\xi_0)}$. It would be more convenient to work with the dimensionless variables $a\equiv \frac{x-\xi_0}{y}$ and $b\equiv\frac{x-x_0}{x_0-\xi_0}$. It is not difficult to check that $P$ can be written in the following form:

\begin{equation}
P(x,y,\xi_0,x_0)= \left( \frac{1+b}{a(1+a^2)}\left(1+(\frac{ab}{1+b})^2\right) \right)^{\frac{1}{3}h_1} \left( \frac{a^3(1+a^2)}{(1+b)^2(1+\frac{ab}{1+b})^2)^3}\right)^{\frac{1}{6}h_\rho}g(a,b)
\label{LPP-CFT4}
\end{equation}
From the above formula, one realizes that all coefficients can be absorbed in $g$ and so $P$ would be a function on $a$ and $b$, i.e. $P(x,y,\xi_0,x_0)=P(\frac{x-\xi_0}{y},\frac{x-x_0}{x_0-\xi_0})$. Now one can apply the Eq [\ref{LPP-SLE(kr)}] to the Eq [\ref{LPP-CFT4}], writing the derivatives in terms of $a$ and $d$, the following differential equation for $P$ is obtained:

\begin{equation}
\left[ \lambda_{a^2}\partial_a^2+\lambda_{b^2}\partial_b^2+\lambda_{ab}\partial_a\partial_b+\lambda_a\partial_a+\lambda_b\partial_b+\lambda \right] h=0
\label{LPPE}
\end{equation}
where:
\begin{equation}
\begin{split}
\lambda_{a^2}&=\frac{\kappa}{2}\\
\lambda_{b^2}&=\frac{\kappa}{2}\left( \frac{b(1+b)}{a}\right) ^2\\
\lambda_{ab}&=-\kappa\frac{b(1+b)}{a}\\
\lambda_{a}&=-\kappa h_{\rho} \frac{1+b}{a}+4\frac{a}{1+a^2} \\
\lambda_{b}&=2(1+b)\left( \frac{1}{1+a^2}-\left( \frac{1+b}{a}\right)^2\right) +\kappa(h_\rho+1)b\left( \frac{1+b}{a}\right)^2\\
\lambda &=\frac{h_{\rho}}{2}\left[\kappa(h_{\rho}+1)-8\right]\left(\frac{1+b}{a}\right)^2\\
\end{split}
\label{def_coef}
\end{equation}
The boundary conditions of the Eq [\ref{LPPE}] are as follows:

\begin{equation}
\begin{split}
&\text{in the region  } x<0  , y\rightarrow 0 \text{ }(a\rightarrow -\infty , b<0)\Rightarrow P\rightarrow 0\\
&\text{in the region  } x>0  , y\rightarrow 0 \text{ }(a\rightarrow +\infty , b<0)\Rightarrow P\rightarrow 1\\
&\text{in the region  } x_0 \rightarrow\infty \text{ }   (b\rightarrow -1) \Rightarrow P\rightarrow P_{\text{chordal}}\\
&\text{in the region  } x_0 \rightarrow 0 \text{     }   (b\rightarrow \pm\infty) \Rightarrow P\rightarrow P_{\text{chordal}}\\
\end{split}
\label{bndry-cond}
\end{equation}
It is notable that in the limit $b\rightarrow -1$, the Eq [\ref{LPPE}] becomes:
\begin{equation}
4a\partial_aP+\frac{\kappa}{2}\left( a^2+1\right) \partial_a^2P=0
\end{equation}
which is exactly the Eq [\ref{EFPF}] in which $a=u|_{x_\infty\rightarrow \infty}$, so the requirement of the last line of Eq [\ref{bndry-cond}] is confirmed.

\section{Results for $\kappa=2$ and $h_{\rho}=0$ at large $x_0$}\label{num}

Since it's hard to solve Eq [\ref{LPPE}] in the general form, this section is devoted to analysis of this equation for the special case $\kappa=2$ and $h_{\rho}=0$ for large $x_0$ ($b\rightarrow-1$). One of the most important examples of this case is the Abelian sandpile model with a sink on the boundary\cite{Ruelle}. In this model the boundary condition changing (bcc) operator corresponding to the change from open to close boundary condition, is the twisting operator $\mu$ with the conformal weight $\frac{-1}{8}$. It has been proved that in the scaling limit, the operator corresponding to a sink on the boundary, results from operator product expansion (OPE) of two twist operators, which is $\tilde{I}=:\bar{\theta}\theta:(z)$ with the conformal weight $0$ in which $\theta$ and $\bar{\theta}$ are grassman variables, living in the ghost action in $c=-2$ CFT. This operator is the logarithmic partner of the identity operator $I$. Let us define $\chi\equiv \frac{y(x_0-\xi_0)}{(x-x_0)^2+y^2}=\frac{a(1+b)}{(1+b)^2+a^2b^2}$. In the limit $x_0\rightarrow\infty$, $\chi$ becomes equal to $\frac{1+b}{a}\equiv\epsilon$ which is a small quantity and we take it as the perturbation parameter. Setting $\kappa=2$ and $h_{\rho}=0$ in Eq [\ref{LPPE}], to first order of $\epsilon$, one obtains:

\begin{equation}
\left[\partial_{a^2}+\epsilon\partial_a\partial_b+\frac{2a\epsilon}{1+a^2}\partial_b+\frac{4a}{1+a^2}\partial_a\right]P=0.
\label{reduced-LPP}
\end{equation}
We expect $P$ not to be singular in the defined domain, so we can expand it in terms of $\chi$. In the first order of $\epsilon$ we have:
\begin{equation}
P=P_0(a)+\frac{y(x_0-\xi_0)}{(x-x_0)^2+y^2} P_1(a)+O(\chi^2)|_{x_0\rightarrow\infty}=P_0(a)+\epsilon P_1(a)+O(\epsilon^2)
\end{equation}
The above ansatz is the only answer satisfying the following conditions; in the limit $x_0\rightarrow\infty$ or $y\rightarrow\infty$ it retrieves the $\rho$-free solution (LPP of SLE($\kappa,\rho=0$)) as expected. Other candidates for the coefficient of $P_1$ are excluded by the similar arguments. Substituting this into Eq [\ref{reduced-LPP}], to the leading order we obtain ($\partial_{b}P_0=\partial_{b}P_0=0$):
\begin{equation}
\partial_{a}^{2}P_0+\frac{4a}{1+a^2}\partial_{a} P_0+\epsilon\left(\partial_{a}^{2}P_1
+\frac{4a}{1+a^2}\partial_{a} P_1-\frac{2}{1+a^2}P_1\right)=0 
\end{equation}
From the above we can conclude that $P_0$ is exactly the solution of Eq [\ref{EFPF}], i.e. Eq [\ref{EFPF2}]. So $P_1$ should be the solution of the following equation:
\begin{equation}
\partial_{a}^{2}P_1+\frac{4a}{1+a^2}\partial_{a} P_1-\frac{2}{1+a^2}P_1=0 
\label{P1}
\end{equation}
with the boundary conditions $\lim_{a\rightarrow\infty} \frac{1}{a}P_1(-\infty)=\lim_{a\rightarrow\infty}\frac{1}{a}P_1(\infty)=0$. The general solution of Eq [\ref{P1}] is:
\begin{equation}
P_1= A{\text{ }}_2F_1\left(\frac{3-\sqrt{17}}{4},\frac{3+\sqrt{17}}{4},\frac{1}{2},-a^2\right) +B {\text{ }}_2F_1\left(\frac{5-\sqrt{17}}{4},\frac{5+\sqrt{17}}{4},\frac{3}{2},-a^2\right)a
\label{P1sol}
\end{equation}
in which $A$ and $B$ are some coefficients to be determined by boundary conditions. The solution Eq [\ref{P1sol}] satisfies the expected boundary conditions. $A$ and $B$ should be determined by the condition $b\rightarrow\infty$ which is beyond our analysis. Therefore to first order of $\epsilon$ we suffice to present the general solution of Eq [\ref{reduced-LPP}]:
\begin{equation}
\begin{split}
P= &\frac{1}{2}+\frac{1}{\sqrt{\pi}\Gamma(\frac{3}{2})} {_2F_1}\left(\frac{1}{2},2,\frac{3}{2},-a^2\right)a+
A(\frac{1+b}{a}){_2F_1}\left(\frac{3-\sqrt{17}}{4},\frac{3+\sqrt{17}}{4},\frac{1}{2},-a^2\right) \\
&+B(1+b) {_2F_1}\left(\frac{5-\sqrt{17}}{4},\frac{5+\sqrt{17}}{4},\frac{3}{2},-a^2\right)
\end{split}
\label{P1sol}
\end{equation}

\section{Conclusion}
In this paper we have calculated the left passage probability for SLE($\kappa,\rho$) for $\rho=\kappa-6$ and general $\rho$. As an example, we have presented the results for the case $\kappa=2$ and $h_{\rho}=0$ at large $x_0$ ($x_0$ is the point on the real axis at which the boundary conditions change). For this case we found the LPP, up to 2 unknown parameters existing in the solution which should be fixed from the general solution including the boundary conditions at $x_0\rightarrow 0$ ($b\rightarrow\infty$ and $-\infty$) keeping $x$ and $y$ fixed.

\begin{acknowledgements}
I wish to thank S. Moghimi-Araghi for useful hints on this work. I also thank M. Ghasemi Nezhadhaghighi for his helpful numerical programming to check some results of the paper.
\end{acknowledgements}

\end{document}